\documentclass{elsart}

\usepackage{amssymb}

\begin{document}

\begin{frontmatter}



\title{Two-state Markovian theory of input-output frequency and phase
synchronization}


\author{Jes\'us Casado-Pascual\corauthref{cor1}}\ead{jcasado@us.es},
\author{Jos\'e  G\'omez-Ord\'o\~nez}, \author{Manuel Morillo}

\address{F\'{\i}sica Te\'orica,
Universidad de Sevilla, Apartado de Correos 1065, Sevilla 41080,
Spain} \corauth[cor1]{Corresponding author, tel: +34-954550945, fax:
+34-954612097.}
\begin{abstract}
A Markovian dichotomic system driven by a deterministic
time-periodic force is analyzed in terms of the statistical
properties of the switching events between the states. The
consideration of the counting process of the switching events leads
us to define a discrete phase. We obtain expressions for the
instantaneous output frequency and phase diffusion associated to the
dichotomic process, as well as for their cycle averages. These
expressions are completely determined by the rates of escape from
both states. They are a convenient starting point for the study of
the stochastic frequency and phase synchronization in a wide range
of situations (both classical and quantum) in which two metastable
states are involved.

\end{abstract}

\begin{keyword}
Synchronization \sep Markovian \sep two-state \sep phase \sep
dichotomic \sep frequency

\PACS 05.40.-a \sep 05.45.Xt \sep 05.10.Gg \sep 02.50.-r

\end{keyword}
\end{frontmatter}

\section{Introduction}
\label{Introduction} The phenomenon of synchronization in stochastic
systems has attracted much interest in recent years
\cite{Freund2,Rozenfeld,Callenbach2002,Freund1,talk03,Lindner,talk04}.
In the case of a periodically driven bistable stochastic system, it
has been shown to be particularly useful to introduce a discrete
phase associated to the output signal
\cite{Freund2,Rozenfeld,talk03,Lindner}. Then, the synchronization
between the periodic input signal and the stochastic output signal
is characterized in terms of certain quantities as, for example, the
average output frequency and the average phase diffusion. Noise
regulates the input-output synchronization in such a way that the
value of the average output frequency might match the value of the
input frequency for a range of noise strength values. Furthermore,
even though the output phase is a stochastic quantity, it is
characterized by a diffusion coefficient with a sharp minimum for
values of the noise strength where frequency locking exists. Noise
induced synchronization is one of the diverse type of phenomena
arising from the interplay between nonlinearity and noise in systems
externally driven by time-periodic forces. Stochastic resonance
\cite{EPL89,PRA91,gammaitoni1998,hanggi2002,casdenk03,casgom03,JCPPRL,casgom04}
and Brownian motors \cite{BM,BM1,BM2,BM3} provide other instances of
this rich phenomenology.

In this paper we focus on the calculation of explicit expressions
for the instantaneous output frequency and phase diffusion, as well
as for their cycle averages. We restrict our study to the case of a
Markovian two-state system, driven by an arbitrary deterministic
time-periodic force, in contrast to Ref~\cite{Freund2}, where only
dichotomic forces are considered. In our analysis, we do not specify
the underlying continuous dynamics which leads, by contraction, to
the two-state description. In this sense, our results can be applied
to the study of the stochastic frequency and phase synchronization
in a wide range of situations (both classical and quantum), whenever
the system presents two metastable states and the contracted
two-state dynamics is (approximately) Markovian. As it has been
shown elsewhere \cite{Jesus}, in the case of a rocked overdamped
bistable stochastic system, such a situation takes place in the
weak-noise and low-frequency limit.

The paper is organized as follows: We begin with a description of
the Markovian two-state model in terms of the switching times and
define the discrete phase associated to this dichotomic process.
Subsequently, in Sect.~\ref{sect3}, we analyze the statistical
properties of the two-state process and the discrete phase. Making
use of the previous results, in Sect.~\ref{sect4} we obtain
expressions for the instantaneous output frequency and phase
diffusion, as well as for their cycle averages. The main conclusions
of this work are presented in Sect.~\ref{sect5}, and some
mathematical details are reported in the Appendix.

\section{The two-state model: switching times and definition of the
  output discrete phase}
\label{sect2}
We consider a two-state Markovian stochastic process
$\chi^{\alpha_0,t_0}(t)$ which only takes the values $+1$ and
$-1$. The notation used emphasizes the fact that at an initial instant
of time $t_0$ the system is in the state $\alpha_0$, i.e.,
$\chi^{\alpha_0,t_0}(t_0)=\alpha_0$ with $\alpha_0=+1$ or $-1$. Due to
some physical mechanism (noise in the case of a classical system, or
the combined action of tunneling and noise in a quantum problem),
there are switches between those two states which are also modulated
by the action of an external, periodic, time-dependent force with
period $T$.  The instant of time at which the $n$-th switch of state
takes place is a random variable which will be denoted by
$\mathcal{T}_{n}^{\alpha_0,t_0}$, with $n=1,2\dots$.

The analysis of the statistics of switching times
$\mathcal{T}_{n}^{\alpha_0,t_0}$, can be carried out along the lines
of the renewal theory detailed by Cox in Ref.~\cite{Cox}, with the
necessary extensions to explicitly time dependent situations. The
random variable $\mathcal{T}_{n}^{\alpha_0,t_0}$ will be
characterized by its probability {\it density} function
\begin{equation}
\label{defgn}
g_n^{\alpha_0,t_0}(t)=\lim_{\Delta t\rightarrow
0^+}\frac{\mathrm{Prob}\left[t<\mathcal{T}_n^{\alpha_0,t_0}\leq
t+\Delta t\right]}{\Delta t},
\end{equation}
or equivalently, by the cumulative distribution function
\begin{equation}
\label{defGn}
G_n^{\alpha_0,t_0}(t)=\mathrm{Prob}\left[\mathcal{T}_n^{\alpha_0,t_0}\leq
t\right]=\int_{t_0}^{t} dt^{\prime} \; g_n^{\alpha_0,t_0}(t^{\prime}),
\end{equation}
or its complementary,
\begin{equation}
\label{defcalGn}
\mathcal{G}_n^{\alpha_0,t_0}(t)=
\mathrm{Prob}\left[\mathcal{T}_n^{\alpha_0,t_0}>
t\right]=1-G_n^{\alpha_0,t_0}(t).
\end{equation}

If we introduce the stochastic process
\begin{equation}
\label{def2}
N^{\alpha_0,t_0}(t)=\max \left[n: \mathcal{T}_{n}^{\alpha_0,t_0}\leq t
\right],
\end{equation}
characterizing the number of switches of state in the interval
$(t_0,t]$, then the two-state stochastic process
$\chi^{\alpha_0,t_0}(t)$ can be expressed as
\begin{equation}
\label{defalpha}
\chi^{\alpha_0,t_0}(t)=\alpha_0 \cos\left[\pi N^{\alpha_0,t_0}(t)  \right].
\end{equation}
Taking into account the above expression, we will define a discrete
phase $\varphi^{\alpha_0, t_0}(t)$ associated to the two-state
stochastic process as
\begin{equation}
\label{defphase}
\varphi^{\alpha_0, t_0}(t)=\pi N^{\alpha_0,t_0}(t).
\end{equation}

\section{Statistical characterization of the discrete phase $\varphi^{\alpha_0, t_0}(t)$ and the
two-state process $\chi^{\alpha_0,t_0}(t)$}
\label{sect3}
The one-time statistical properties of the discrete phase
$\varphi^{\alpha_0,t_0}(t)$ can be evaluated from the probability
distribution of the number of switches of state
\begin{equation}
\label{defpn}
\rho_n^{\alpha_0,t_0}(t)=\mathrm{Prob}\left[N^{\alpha_0,t_0}(t)=n
\right],
\end{equation}
with $n=0,1,2,\dots$. Making use of the definition of
$N^{\alpha_0,t_0}(t)$ in Eq.~(\ref{def2}) it is easy to see that
\begin{equation}
\label{prop1}
\mathrm{Prob}\left[N^{\alpha_0,t_0}(t)\geq n
\right]=G_n^{\alpha_0,t_0}(t),
\end{equation}
with $G_0^{\alpha_0,t_0}(t)=1$. Consequently, the probability
distribution of the number of switches of state and its derivative with
respect to $t$ can be expressed, respectively, as
\begin{eqnarray}
\label{pn1}
\rho_n^{\alpha_0,t_0}(t)&=&G_n^{\alpha_0,t_0}(t)-G_{n+1}^{\alpha_0,t_0}(t)
\nonumber\\
&=&\mathcal{G}_{n+1}^{\alpha_0,t_0}(t)-\mathcal{G}_{n}^{\alpha_0,t_0}(t),
\end{eqnarray}
and
\begin{equation}
\label{meq1}
\dot{\rho}_n^{\alpha_0,t_0}(t)=g_n^{\alpha_0,t_0}(t)-g_{n+1}^{\alpha_0,t_0}(t),
\end{equation}
with $\mathcal{G}_0^{\alpha_0,t_0}(t)=g_0^{\alpha_0,t_0}(t)=0$.

In order to express the above equation in a more transparent form, it
is convenient to introduce the probability of an almost immediate
switch of state after $n$ switches
\begin{equation}
\label{defrate1}
\Gamma_{n}^{\alpha_0,t_0}(t)=\lim_{\Delta t\rightarrow 0^{+}}
\frac{\mathrm{Prob}\left[t<\mathcal{T}_{n+1}^{\alpha_0,t_0}
\leq t+\Delta t\left| \right.N^{\alpha_0,t_0}(t)=n
\right]}{\Delta t},
\end{equation}
as well as the probability of an almost immediate switch of state from
state $\beta$
\begin{equation}
\label{defrate3}
\gamma_{\beta}^{\alpha_0,t_0}(t)=\lim_{\Delta t\rightarrow 0^{+}}
\frac{\mathrm{Prob}\left[t<\mathcal{T}_{N^{\alpha_0,t_0}(t)+1}^{\alpha_0,t_0}
\leq t+\Delta t\left| \right. \chi^{\alpha_0,t_0}(t)=\beta
\right]}{\Delta t}.
\end{equation}
Due to the Markovian character of the stochastic process
$\chi^{\alpha_0,t_0}(t)$, it is clear that the rate
$\gamma_{\beta}^{\alpha_0,t_0}(t)$ is independent of the initial
preparation $\alpha_0$ at $t_0$ and, for this reason, henceforth it
will be denoted by $\gamma_{\beta}(t)$. Analogously, the rate
$\Gamma_{n}^{\alpha_0,t_0}(t)$ only depends on the value of
$\chi^{\alpha_0,t_0}(t)$ after the $n$-th switch of state. Thus,
noting that after an even number of switches of state the system ends
up in the same state as it was initially, whereas for an odd number of
switches the system ends up in the other state, it follows that
\begin{equation}
\label{defrate4}
\Gamma_{n}^{\alpha_0,t_0}(t)=\gamma_{\alpha_n}(t),
\end{equation}
where $\alpha_n=(-1)^n \alpha_0$. The time-periodicity of the external
force becomes manifest in the fact that the rate $\gamma_{\beta}(t)$
is also a periodic function of time with the same period $T$, i.e.,
\begin{equation}
\label{periodicity}
\gamma_{\beta}(t+T)=\gamma_{\beta}(t).
\end{equation}

Multiplying and dividing the right-hand side of Eq.~(\ref{defrate1})
by $\rho_n^{\alpha_0,t_0}(t)$, assuming that the probability of more
than one switch of state between $t$ and $t+\Delta t$ is
$\mathcal{O}\left[(\Delta t)^2\right]$, and taking into account
Eqs.~(\ref{defgn}) and (\ref{defrate4}), it is easy to see that
\begin{equation}
\label{defrate2}
\gamma_{\alpha_n}(t)=\frac{g_{n+1}^{\alpha_0,t_0}(t)}{\rho_n^{\alpha_0,t_0}(t)}.
\end{equation}
Then, from Eq.~(\ref{meq1}) one obtains the following hierarchy of
differential equations for $\rho_n^{\alpha_0,t_0}(t)$
\begin{eqnarray}
\label{meq2}
\dot{\rho}_n^{\alpha_0,t_0}(t)&=&-\gamma_{\alpha_n}(t)\rho_{n}^{\alpha_0,t_0}(t)
+\gamma_{\alpha_{n-1}}(t)\rho_{n-1}^{\alpha_0,t_0}(t)\, \mbox{ for }n\geq 1,\\
\dot{\rho}_0^{\alpha_0,t_0}(t)&=&-\gamma_{\alpha_0}(t)\rho_{0}^{\alpha_0,t_0}(t),
\end{eqnarray}
which must be solved with the initial condition
$\rho_{n}^{\alpha_0,t_0}(t_0)=\delta_{n,0}$.

We will also introduce the probability of $\chi^{\alpha_0,t_0}(t)$ to
take on the value $\beta$
\begin{eqnarray}
\label{defp}
p_{\beta}(t|\alpha_0,t_0)&=&\mathrm{Prob}\left[\chi^{\alpha_0,t_0}(t)=\beta\right]\nonumber\\
&=&\delta_{\alpha_0,\beta}\sum_{n=0}^{\infty}\rho_{2
n}^{\alpha_0,t_0}(t)+\delta_{-\alpha_0,\beta} \sum_{n=0}^{\infty}\rho_{2
n+1}^{\alpha_0,t_0}(t),
\end{eqnarray}
with $\beta=1 \mbox{ or }-1$. Notice that in the notation we have made
explicit the conditional dependence of this probability on the initial
preparation.

Differentiating the above expression with respect to $t$
and taking into account Eq.~(\ref{meq2}), it is straightforward to
obtain the master equation for the Markovian stochastic process
$\chi^{\alpha_0,t_0}(t)$
\begin{equation}
\label{masterequation1}
\dot{p}_{\beta}(t|\alpha_0,t_0)=-\gamma_{\beta}(t)
p_{\beta}(t|\alpha_0,t_0)+\gamma_{-\beta}(t)
p_{-\beta}(t|\alpha_0,t_0),
\end{equation}
which must be solved with the initial condition
$p_{\beta}(t_0|\alpha_0,t_0)=\delta_{\alpha_0,\beta}$. The solution of
the above equation can be easily obtained taking into account that
$p_{-\beta}(t|\alpha_0,t_0)=1-p_{\beta}(t|\alpha_0,t_0)$, and the
result is
\begin{eqnarray}
\label{sol1} p_{\beta}(t|\alpha_0,t_0)&=&\delta_{\alpha_0,\beta}
\exp\left[-\int_{t_0}^{t} d t'\, \gamma(t')\right]\nonumber \\&&+
\int_{t_0}^{t}d t'\,\gamma_{-\beta}(t')\exp\left[-\int_{t'}^{t} d
t''\, \gamma(t'')\right],
\end{eqnarray}
where
\begin{equation}
\label{gammdef}
\gamma(t)=\gamma_{+1}(t)+\gamma_{-1}(t).
\end{equation}
In order to obtain an expression independent of the initial
preparation, it is necessary to take the limit $t_0\rightarrow
-\infty$ of Eq.~(\ref{sol1}) and introduce the long-time populations
\begin{equation}
\label{sol2} p_{\beta}(t)=\lim_{t_0\rightarrow -\infty}
p_{\beta}(t|\alpha_0,t_0)=\int_{-\infty}^{t}d
t'\,\gamma_{-\beta}(t')\,\exp\left[-\int_{t'}^{t} d t''\,
\gamma(t'')\right].
\end{equation}
From the results contained in Appendix~\ref{app1}, it follows that the
long-time population $p_{\beta}(t)$ is a periodic function of $t$ with
period $T$ which can be expressed, for $t\in[0,T]$, as
\begin{eqnarray}
\label{longtimepop} p_{\beta}(t)&=&\frac{1}{2}
\mathrm{csch}\left(\frac{\bar{\gamma} T}{2}\right)\int_{0}^T d
t'\,\gamma_{-\beta}(t')\nonumber\\&& \times\exp\left[\mathrm{sgn}
(t-t')\frac{\bar{\gamma} T}{2} -\int_{t'}^{t} d t''\,
\gamma(t'')\right],
\end{eqnarray}
with $\bar{\gamma}$ given by Eq.~(\ref{appeq5}).

\section{The output frequency and the phase diffusion}
\label{sect4} The instantaneous output frequency and phase diffusion
are, respectively, defined as \cite{Freund2}
\begin{equation}
\label{defoutputfreq}
\Omega_\mathrm{out}^{\alpha_0,t_0}(t)=\frac{\partial}{\partial t}\left\langle
\varphi^{\alpha_0,t_0}(t)\right\rangle=\pi \frac{\partial}{\partial t}\left\langle
N^{\alpha_0,t_0}(t)\right\rangle
\end{equation}
and
\begin{eqnarray}
\label{phasedispdef1}
D_\mathrm{out}^{\alpha_0,t_0}(t)&=&\frac{\partial}{\partial t}
 \left\{\left\langle \left[
 \varphi^{\alpha_0,t_0}(t)\right]^2\right\rangle-\left\langle
 \varphi^{\alpha_0,t_0}(t)\right\rangle^2\right\}\nonumber\\ &=&\pi^2
 \frac{\partial}{\partial t} \left\{\left\langle \left[
 N^{\alpha_0,t_0}(t)\right]^2\right\rangle-\left\langle
 N^{\alpha_0,t_0}(t)\right\rangle^2\right\},
\end{eqnarray}
where $\langle \dots \rangle$ denotes the average with respect to
the distribution $\rho_{n}^{\alpha_0,t_0}(t)$, i.e.,
\begin{equation}
\label{average}
\left\langle K\left[N^{\alpha_0,t_0}(t) \right]\right\rangle =\sum_{n=0}^{\infty}
K(n)\rho_n^{\alpha_0,t_0}(t),
\end{equation}
$K\left[N^{\alpha_0,t_0}(t)\right]$ being an arbitrary one-time
function of $N^{\alpha_0,t_0}(t)$.

Multiplying Eq.~(\ref{meq2}) by $n$, summing up the series
$\sum_{n=1}^{\infty}n \dot{\rho}_n^{\alpha_0,t_0}(t)$, and taking into
account Eq.~(\ref{defp}), it is easy to obtain that
\begin{equation}
\label{freqoutres}
\Omega_\mathrm{out}^{\alpha_0,t_0}(t)=\pi \left[\gamma_{+1}(t)
p_{+1}(t|\alpha_0,t_0)+\gamma_{-1}(t) p_{-1}(t|\alpha_0,t_0)\right].
\end{equation}

Analogously, if one sums up the series $\sum_{n=1}^{\infty}n^2
\dot{\rho}_n^{\alpha_0,t_0}(t)$ by using Eq.~(\ref{meq2}), it results
after some simplifications that
\begin{equation}
\label{dif1}
\frac{\partial}{\partial t} \left\langle \left[
 N^{\alpha_0,t_0}(t)\right]^2\right\rangle= \pi^{-1}
 \Omega_\mathrm{out}^{\alpha_0,t_0}(t)
 +2\sum_{\beta=\pm \alpha_0} \gamma_{\beta}(t) \Phi_{\beta}^{\alpha_0,t_0}(t),
\end{equation}
where
\begin{equation}
\label{dif23}
\Phi_{\beta}^{\alpha_0,t_0}(t)=\delta_{\beta,\alpha_0}
 \sum_{n=0}^{\infty}2n \,\rho_{2
 n}^{\alpha_0,t_0}(t)+\delta_{\beta,-\alpha_0}\sum_{n=0}^{\infty}(2n+1)
 \rho_{2 n+1}^{\alpha_0,t_0}(t).
\end{equation}
Replacing Eq.~(\ref{dif1}) into Eq.~(\ref{phasedispdef1}) and taking
into account Eqs.~(\ref{defoutputfreq}), (\ref{freqoutres}), and
(\ref{dif23}), it is straightforward to see that
\begin{equation}
\label{phasedis3}
D_\mathrm{out}^{\alpha_0,t_0}(t)=\pi\,
\Omega_\mathrm{out}^{\alpha_0,t_0}(t)+2\pi^2\Delta\gamma(t)\Psi^{\alpha_0,t_0}(t),
\end{equation}
with
\begin{equation}
\label{defdgamma}
\Delta\gamma(t)=\gamma_{+1}(t)-\gamma_{-1}(t),
\end{equation}
and
\begin{equation}
\label{Psidef}
\Psi^{\alpha_0,t_0}(t)=\alpha_0
\left[\Phi_{\alpha_0}^{\alpha_0,t_0}(t)p_{-\alpha_0}(t|\alpha_0,t_0)-
\Phi_{-\alpha_0}^{\alpha_0,t_0}(t)p_{\alpha_0}(t|\alpha_0,t_0)
\right].
\end{equation}
Making use of Eqs.~(\ref{meq2}) and (\ref{masterequation1}) and after
some lengthy calculations, it is possible to prove that
$\Psi^{\alpha_0,t_0}(t)$ satisfies the differential equation
\begin{equation}
\label{phasedis4}
\dot{\Psi}^{\alpha_0,t_0}(t)=-\gamma(t)
\Psi^{\alpha_0,t_0}(t)-\sum_{\beta=\pm 1}\beta \,\gamma_{\beta}(t)
\left[p_{\beta}(t|\alpha_0,t_0)\right]^2,
\end{equation}
with the initial condition $\Psi^{\alpha_0,t_0}(t_0)=0$. The above
equation can be formally solved, yielding
\begin{equation}
\label{phasedis6} \Psi^{\alpha_0,t_0}(t)=-\sum_{\beta=\pm 1} \beta
\int_{t_0}^{t} d t^{\prime} \,\gamma_{\beta}(t^{\prime})\left[
p_{\beta}(t^{\prime}|\alpha_0,t_0)\right]^2
\exp\left[-\int_{t^{\prime}}^{t} d t^{\prime \prime}
\gamma(t^{\prime\prime})\right].
\end{equation}
Inserting Eq.~(\ref{phasedis6}) into Eq.~(\ref{phasedis3}), one
obtains
\begin{eqnarray}
\label{phasedis7} D_\mathrm{out}^{\alpha_0,t_0}(t)&=&\pi\,
\Omega_\mathrm{out}^{\alpha_0,t_0}(t)-2\pi^2\Delta\gamma(t)
\nonumber\\&&\times\sum_{\beta=\pm 1} \beta \int_{t_0}^{t} d
t^{\prime} \gamma_{\beta}(t^{\prime})
\left[p_{\beta}(t^{\prime}|\alpha_0,t_0)\right]^2
\exp\left[-\int_{t^{\prime}}^{t} d t^{\prime \prime}
\gamma(t^{\prime\prime})\right].
\end{eqnarray}
Equations~(\ref{freqoutres}) and (\ref{phasedis7}) for the
instantaneous output frequency and phase diffusion, respectively,
still depend on the initial preparation. In order to obtain
expressions independent on the initial preparation, it is necessary
to take the limit $t_0\rightarrow-\infty$. In this limit, one
obtains
\begin{equation}
\label{freqoutreslongtime}
\Omega_\mathrm{out}(t)=\lim_{t_0\rightarrow-\infty}
\Omega_\mathrm{out}^{\alpha_0,t_0}(t)=\pi \left[\gamma_{+1}(t)
p_{+1}(t)+\gamma_{-1}(t) p_{-1}(t)\right]
\end{equation}
and
\begin{eqnarray}
\label{phasedis7longtime}
D_\mathrm{out}(t)&=&\lim_{t_0\rightarrow
  -\infty}D_\mathrm{out}^{\alpha_0,t_0}(t)=\pi\,
\Omega_\mathrm{out}(t)-2\pi^2\Delta\gamma(t)\nonumber\\&&\times
\sum_{\beta=\pm 1} \beta \int_{-\infty}^{t} d t^{\prime}
\gamma_{\beta}(t^{\prime}) \left[p_{\beta}(t^{\prime})\right]^2
\exp\left[-\int_{t^{\prime}}^{t} d
  t^{\prime \prime}
  \gamma(t^{\prime\prime})\right].\end{eqnarray}

The functions $\Omega_\mathrm{out}(t)$ and $D_\mathrm{out}(t)$ are
 periodic functions of the time $t$ (see Appendix~\ref{app1}). Thus,
 one can perform a cycle average and define the average output
 frequency
\begin{equation}
\label{finaloutfreq}
\Omega_\mathrm{out}=\frac{1}{T}\int_{0}^{T}
dt\,\Omega_\mathrm{out}(t)=\frac{\pi}{T} \int_{0}^{T} dt
\,\left[\gamma_{+1}(t) p_{+1}(t)+\gamma_{-1}(t) p_{-1}(t)\right],
\end{equation}
and the average phase diffusion
\begin{eqnarray}
\label{phasediff2}
D_\mathrm{out}&=&\frac{1}{T}\int_{0}^{T} dt\, D_\mathrm{out}(t)=\pi\,
\Omega_\mathrm{out}
-\frac{\pi^2}{T}\mathrm{csch}\left(\frac{\bar{\gamma}
  T}{2}\right)\sum_{\beta=\pm 1} \beta
\int_{0}^{T}d t\int_{0}^{T} d t^{\prime} \Delta\gamma(t)
\nonumber\\&&\times\gamma_{\beta}(t^{\prime})\left[p_{\beta}(t^{\prime})\right]^2
\exp\left[\mathrm{sgn} (t-t^{\prime})\frac{{\bar{\gamma}}
    T}{2}-\int_{t^{\prime}}^{t} d t^{\prime \prime}
  \gamma(t^{\prime\prime})\right],
\end{eqnarray}
where $\bar{\gamma}$ is given by Eq.~(\ref{appeq5}) and we have made
use of Eq.~(\ref{appeq6}) with $h(t)=\gamma_{\beta}(t)
\left[p_{\beta}(t)\right]^2$.
Equations~(\ref{freqoutreslongtime}), (\ref{phasedis7longtime}),
(\ref{finaloutfreq}), and (\ref{phasediff2}), complemented by
Eq.~(\ref{longtimepop}), are the main results of this work.
\section{Concluding remarks}
\label{sect5} In this paper, we have put forward a Markovian
two-state theory to describe the phenomenon of frequency and phase
synchronization in noisy systems driven by deterministic
time-periodic forces. The fundamentals of our strategy are based on
an extension of the renewal theory to time-dependent situations. We
have obtained explicit expressions for the instantaneous output
frequency and phase diffusion, as well as for their cycle averages.
These expressions, which are completely determined by the rates of
escape from both states, are quite general and should apply to
realistic situations with two metastable configurations. Thus, they
represent a convenient starting point for the study of the frequency
and phase synchronization in a wide range of situations.

\section*{Acknowledgements}
We acknowledge the support of the Direcci\'on General de Ense\~nanza
  Superior of Spain (BFM2002-03822) and the Junta de
  Andaluc\'{\i}a. We also want to gratefully acknowledge Prof. Peter
  H\"anggi for suggesting the problem and for fruitful discussions.

\appendix

\section{Some mathematical details}
\label{app1}
The aim of this Appendix is to analyze some interesting properties of
integrals of the form
\begin{equation}
\label{appeq1} H(t)=\int_{-\infty}^t d t'\, h(t')
\,\exp\left[-\int_{t'}^{t} d t''\, \gamma(t'')\right],
\end{equation}
with $h(t)$ being a periodic function of $t$ with period $T$. Examples
of integral of this type are those appearing in Eqs.~(\ref{sol2}) and
(\ref{phasedis7longtime}).

First, we will prove that $H(t)$ is also a periodic function of $t$
with the same period $T$ as the functions $h(t)$ and $\gamma(t)$. In
order to do so, let us consider its value at $t+T$,
\begin{equation}
\label{appeq2} H(t+T)=\int_{-\infty}^{t+T} d t'\, h(t')
\,\exp\left[-\int_{t'}^{t+T} d t''\, \gamma(t'')\right].
\end{equation}
After making the change of variables $\tilde{t}'=t'-T$ and
$\tilde{t}''=t''-T$ and using the periodicity of the functions $h(t)$
and $\gamma(t)$, it is straightforward to see that
$H(t+T)=H(t)$.

Once that the periodicity of $H(t)$ has been proved, it is possible to
express $H(t)$ in a more compact form which does not involve an
improper integral. To do so, let us split the integral in
Eq.~(\ref{appeq1}) into the two pieces
\begin{equation}
\label{appeq3} H(t)=H(0) \exp\left[-\int_{0}^{t} d t'\,
\gamma(t')\right] +\int_{0}^t d t'\, h(t') \,
\exp\left[-\int_{t'}^{t} d t''\, \gamma(t'')\right].
\end{equation}
Setting $t=T$ in the above expression and taking into account that
$H(T)=H(0)$, it follows that
\begin{equation}
\label{appeq4} H(0)=\frac{e^{-\bar{\gamma} T}}{1-e^{-\bar{\gamma}
T}} \int_{0}^T d t'\, h(t') \,\exp\left[\int_{0}^{t'} d t''\,
\gamma(t'')\right],
\end{equation}
with
\begin{equation}
\label{appeq5}
\bar{\gamma}=\frac{1}{T} \int_{0}^T d t\,\gamma(t).
\end{equation}
Then, inserting Eq.~(\ref{appeq4}) into Eq.~(\ref{appeq3}), one
obtains after some simplifications that
\begin{equation}
\label{appeq6} H(t)=\frac{1}{2}
\mathrm{csch}\left(\frac{\bar{\gamma} T}{2}\right)\int_{0}^T d t'\,
h(t') \,\exp\left[\mathrm{sgn} (t-t')\frac{\bar{\gamma} T}{2}
-\int_{t'}^{t} d t''\, \gamma(t'')\right],
\end{equation}
for $t\in [0,T]$.

\end{document}